\documentclass[aps,prd,floats,twocolumn,epsf,nofootinbib,showpacs]{revtex4}
\usepackage{latexsym}
\usepackage{amssymb}
\usepackage[dvips]{graphicx}
\usepackage[applemac]{inputenc}
\usepackage[nottoc]{tocbibind} 
\usepackage{float}
\usepackage{fancyhdr}
\usepackage{amsfonts}
\usepackage{amsmath}
\usepackage{color}
\usepackage{mathrsfs}
\usepackage{bm}
\usepackage{epsfig}

\newcommand{\be}{\begin{equation}}
\newcommand{\ee}{\end{equation}}

\def\lsim{\mathrel{\raise.3ex\hbox{$<$\kern-.75em\lower1ex\hbox{$\sim$}}}}

\def\gsim{\mathrel{\raise.3ex\hbox{$>$\kern-.75em\lower1ex\hbox{$\sim$}}}}

\def\beq{\begin{eqnarray}}
\def\eeq{\end{eqnarray}}
\def\bea{\begin{eqnarray}}
\def\eea{\end{eqnarray}}

\begin{document}

\title{Revisiting XENON100's Constraints (and Signals?) For Low-Mass Dark Matter}

\author{Dan Hooper}

\address{
Center for Particle Astrophysics, Fermi National Accelerator
Laboratory, Batavia, IL 60510 \\ 
Department of Astronomy and Astrophysics, University of Chicago,
Chicago, IL 60637}

\begin{abstract}

Although observations made with the CoGeNT and CDMS experiments have been interpreted as possible signals of low-mass ($\sim$7-10 GeV) dark matter particles, constraints from the XENON100 collaboration appear to be incompatible with this hypothesis, at least at face value. In this paper, we revisit XENON100's constraint on dark matter in this mass range, and consider how various uncertainties and assumptions made might alter this conclusion. We also note that while XENON100's two nuclear recoil candidates each exhibit very low ratios of ionization-to-scintillation signals, making them difficult to attribute to known electronic or neutron backgrounds, they are consistent with originating from dark matter particles in the mass range favored by CoGeNT and CDMS. We argue that with lower, but not implausible, values for the relative scintillation efficiency of liquid xenon ($L_{\rm eff}$), and the suppression of the scintillation signal in liquid xenon at XENON100's electric field ($S_{\rm nr}$), these two events could consistently arise from dark matter particles with a mass and cross section in the range favored by CoGeNT and CDMS. If this interpretation is correct, we predict that the LUX experiment, with a significantly higher light yield than XENON100, should observe dark matter induced events at an observable rate of $\sim$3-24 per month.

\end{abstract}

\pacs{95.36.+x; FERMILAB-PUB-13-188-A}

\maketitle


\section{Introduction}

Over the past few years, a number of direct detection experiments have presented results that can be interpreted as signals of low-mass dark matter particles. In particular, the CoGeNT collaboration has reported an excess of low-energy events~\cite{cogent} which, after accounting for surface events and other backgrounds, favors a dark matter particle with a mass of $m_{\rm DM}\simeq6.5$-10 GeV and a spin-independent elastic scattering cross section with nucleons of $\sigma_{\rm SI}\simeq(2-6)\times 10^{-41}$ cm$^2$~\cite{Aalseth:2012if,Kelso:2011gd}. The CoGeNT collaboration has also reported possible evidence for an annual variation in their rate, although with only modest statistical significance~\cite{Aalseth:2011wp} (see also Refs.~\cite{Hooper:2011hd,Fox:2011px,Farina:2011pw,Schwetz:2011xm,Arina:2011zh}). 

More recently, the CDMS collaboration has reinvigorated interest in the CoGeNT excess by reporting the observation of three nuclear recoil-like events in their silicon detectors. The silicon analysis of the CDMS collaboration favors a dark matter interpretation over known-backgrounds at the 99.81\% confidence level (corresponding to slightly more than 3$\sigma$ significance), with a best fit corresponding to $m_{\rm DM}=8.6$ GeV, $\sigma_{\rm SI}=1.9\times 10^{-41}$ cm$^2$~\cite{Agnese:2013rvf}, very near the values implied by CoGeNT. CDMS' silicon events represent the first instance in which a ``zero-background'' experiment has reported a statistically significant excess of events that could be possibly interpreted as a signal of dark matter. 

At very low energies, the CDMS experiment is no longer background-free. Although no low-energy ($\lsim$~7 keV) silicon data has been presented, the CDMS collaboration has published an analysis of their low-energy germanium detectors~\cite{Ahmed:2010wy}. Citing challenges in distinguishing low-energy nuclear recoil events from surface events, zero-charge events, and electron recoil events, the CDMS collaboration has simply derived a conservative upper limit on the dark matter's elastic scattering cross section, assuming that all of the events in their nuclear recoil band are nuclear recoils. The CDMS collaboration has suggested, however, that most of these events could be plausibly attributed to their zero-charge background~\cite{Ahmed:2010wy}. 

Making use of this same data set, Collar and Fields performed an independent likelihood analysis of CDMS' low-energy germanium events~\cite{Collar:2012ed}. Using the high-energy ($\gsim 6$ keV) component of the zero-charge event population to constrain the distribution of such events at lower energies, they found that low-energy zero-charge events could not account for the rate observed in the nuclear recoil band.\footnote{The central point of the analysis in Ref.~\cite{Collar:2012ed} is that CDMS' zero-charge events with recoil energies greater than $\sim6$ keV appear narrowly centered around ionization energies of $\simeq 0.25$ keVee, while the distribution of the lower energy events ($\sim$$3$-5 keV) is centered around a higher ionization energy of $\simeq 0.5$ keVee. As the central value of the zero-charge band's ionization energy is not expected to vary with recoil energy, it is difficult to interpret the low-energy events in question as being dominated by the zero-charge background. The CDMS collaboration is currently in the progress of performing two similar likelihood analyses, making use the same data set, and using new low-background data from SuperCDMS detectors.} Instead, they find a strong statistical preference (5.7$\sigma$) for a nuclear recoil component (relative to the background model). Interpreted as dark matter scattering, this event population favors a mass of $m_{\rm DM}\simeq 7$-10 GeV and an elastic scattering cross section of $\sigma_{\rm SI}\simeq(0.6-6)\times 10^{-41}$ cm$^2$~\cite{Collar:2012ed}.

\begin{figure}[t]
\centering
{\includegraphics[angle=0.0,width=3.5in]{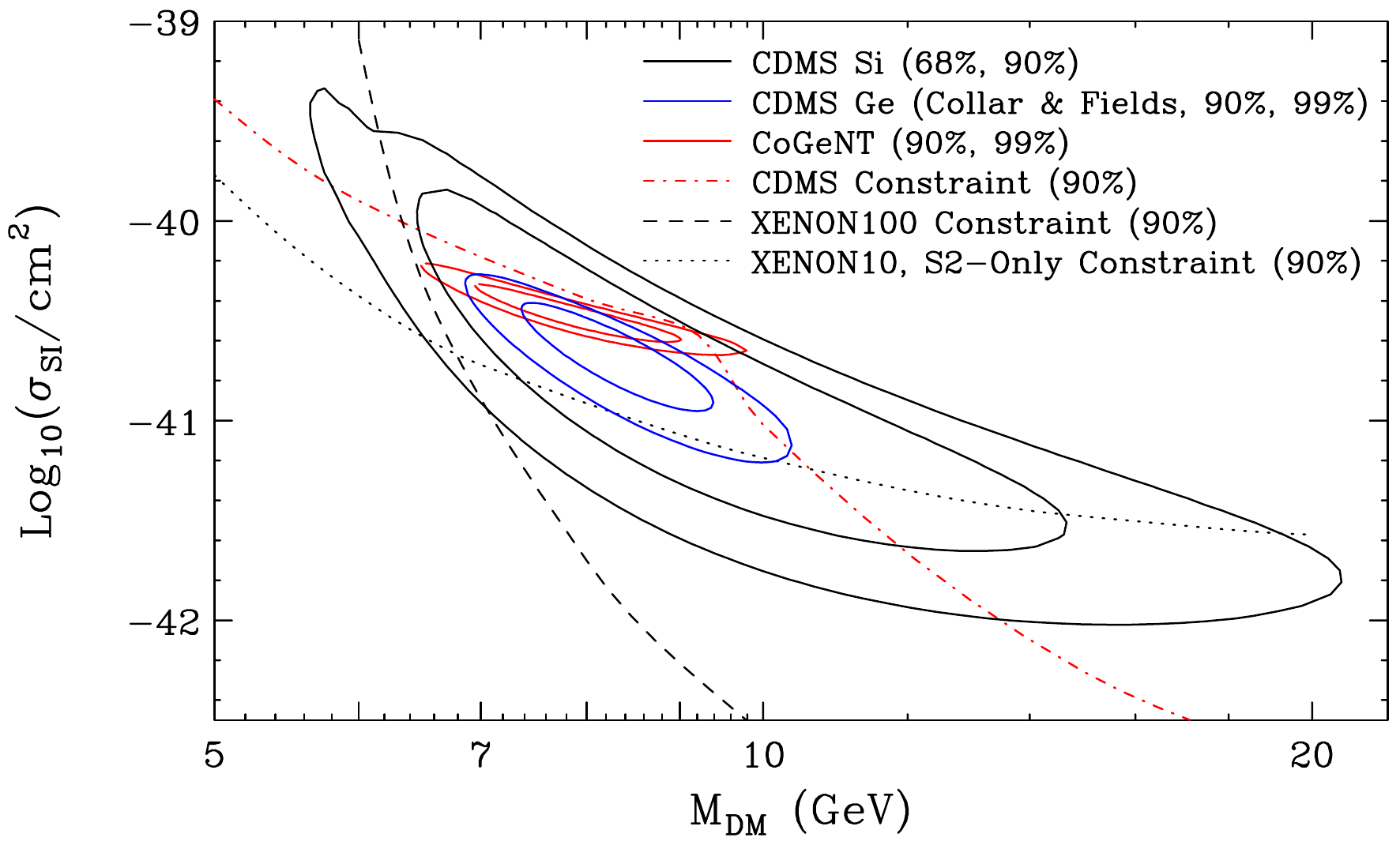}}
\caption{Regions of dark matter parameter space favored by CoGeNT~\cite{Aalseth:2012if}, CDMS (silicon)~\cite{Agnese:2013rvf}, and by Collar and Fields' independent analysis of low-energy CDMS (germanium) data~\cite{Collar:2012ed}. Also shown are the published constraints from CDMS~\cite{Ahmed:2010wy,cdms}, XENON10 (S2-only)~\cite{Angle:2011th}, and XENON100~\cite{Aprile:2012nq}. While there is encouraging overlap between the regions favored by CoGeNT and CDMS, there is a degree of conflict between a dark matter interpretation of these signals and the constraints as presented by the XENON100 and XENON10 collaborations.}
\label{situation}
\end{figure}

In Fig.~\ref{situation}, we show the regions of the dark matter parameter space that are favored by these three potential signals. Encouragingly, they are each consistent with arising from a $m_{\rm DM}\simeq 7$-10 GeV particle, scattering with a spin-independent elastic scattering cross section of $\sigma_{\rm SI}\simeq (2-5)\times 10^{-41}$ cm$^2$. This mass range is also in good agreement with that favored by gamma-ray signals reported from the Galactic Center~\cite{gc1,gc2,gc3,gc4}, and from the much larger region of the Inner Galaxy~\cite{gc5}. While we also note that this mass and elastic scattering cross section is not very far from the regions of parameter space favored by the CRESST-II~\cite{Angloher:2011uu} (see also, Ref.~\cite{Kopp:2011yr}) and DAMA/LIBRA~\cite{damanew} collaborations, we consider these signals to be more difficult to interpret at this time. In particular, although the analysis of the CRESST II collaboration favors a population of dark matter-like events over known backgrounds at greater than $4\sigma$ significance, they only attribute 17-38 of the 67 events in their low-energy nuclear recoil acceptance region to dark matter recoils.  With this modest ratio of signal-to-background, one could reasonably expect uncertainties in the background model to significantly impact the preferred regions of dark matter parameter space (see, for example, Ref.~\cite{Kuzniak:2012zm}). We enthusiastically await results from CRESST's new run, scheduled to begin later this year, which is designed to achieve lower rates of alpha and lead recoil backgrounds. Regarding the annual modulation observed by DAMA/LIBRA, it appears that this signal can be reconciled with the null results of other experiments only if the local dark matter velocity distribution contains a very significant non-isotropic component, such as that associated with tidal streams~\cite{Kelso:2011gd,Purcell:2012sh,tidal1,ls,Freese:2012xd}. In such a scenario, however, it is non-trivial to translate a modulation signal into a corresponding region of the dark matter mass-cross section plane, making it difficult to compare to other signals and constraints.

Also shown in Fig.~\ref{situation} are the most stringent current constraints on the dark matter's elastic scattering cross section in the mass range of interest. The constraints from the CDMS collaboration~\cite{Ahmed:2010wy,cdms} are consistent with the bulk of the parameter space favored by the CoGeNT and CDMS signals. Until recently, the XENON10 collaboration's S2-only analysis (which lowers their threshold by considering events without a scintillation signal) was believed to yield the most stringent constraint for low-mass dark matter particles. An error in the derivation of this limit, however, was recently identified by the authors of Ref.~\cite{Frandsen:2013cna}, and has since been corrected by the XENON10 collaboration (see the erratum to Ref.~\cite{Angle:2011th}), relaxing the original constraint by a factor of 4 at 10 GeV. We note that the derivation of XENON10's constraint as derived in by the authors of Ref.~\cite{Frandsen:2013cna} is still significantly weaker than that shown in the erratum of the XENON10 collaboration's paper. In any case, given the large uncertainties associated with the charge yield of liquid xenon, it is not at all difficult to imagine that this constraint could be even weaker than currently presented (see, for example, Figs.~5 and 6 of Ref.~\cite{Bezrukov:2010qa}).

In light of the revision of XENON10's constraint (and uncertainties associated with the charge yield), the constraint from the XENON100 collaboration~\cite{Aprile:2012nq} now appears to be the only serious obstacle to interpreting the CoGeNT and CDMS signals as evidence of dark matter. If we were only trying to reconcile the XENON100 constraint with a dark matter interpretation of events on either silicon {\it or} germanium, one could invoke carefully chosen isospin violating couplings~\cite{zurek,Feng:2011vu}. For example, if we set the ratio of the dark matter's couplings to proton and neutrons to $f_p/f_n \simeq -0.7$, the rates in xenon-based experiments can be suppressed by more than an order of magnitude relative to those in germanium~\cite{Feng:2011vu}. This choice of $f_p/f_n$, however, also enhances the scattering rate with silicon by a factor of $\sim$7-10, destroying the compatibility of the silicon and germanium signals shown in Fig.~\ref{situation}. Requiring that the ratio of scattering rates with silicon and germanium does not change relative to the standard case ($f_p/f_n=1$) by more than a factor of $\sim$3, we find that isospin violation can suppress the XENON100 constraint, but not by more than a factor of $\sim$2-3, which is insufficient to reconcile it with a dark matter interpretation of the CoGeNT and CDMS signals.

If the signals observed by CoGeNT and CDMS are in fact from the elastic scattering of dark matter, it appears that XENON100's sensitivity to low-energy nuclear recoils must be lower than previously presented. In this paper, we revisit the results of the XENON100 experiment, focusing on their implications for dark matter particles in the mass range collectively favored by CoGeNT and CDMS, $m_{\rm DM}\simeq$7-10 GeV. We note, as previously shown in Refs.~\cite{Sorensen:2012ts,Aprile:2013teh}, that XENON100's two nuclear recoil candidate events exhibit scintillation and ionization signals which are consistent with that predicted to result from the elastic scattering of low-mass dark matter particles, but exhibit far less ionization than would be expected from electronic or neutron background events. Under the same assumptions as made by the XENON100 collaboration, however, the cross section required to account for these two events is two orders of magnitude lower than that implied by the CoGeNT and CDMS signals. If we adopt a lower value for the scintillation efficiency of liquid xenon, and account for the possibility of energy dependence in the suppression of the scintillation signal resulting from the experiment's electric field, we find that it is possible that XENON100's two events could have arisen from the same dark matter species responsible for the excesses observed by CoGeNT and CDMS. Other factors, such as the details of the treatment of scintillation fluctuations and uncertainties in the dark matter velocity distribution, could also help to alleviate the apparent tension between these experiments.

\section{Detecting Low-Mass Dark Matter with XENON100}
\label{setup}

\begin{figure}[t]
\centering
{\includegraphics[angle=0.0,width=3.5in]{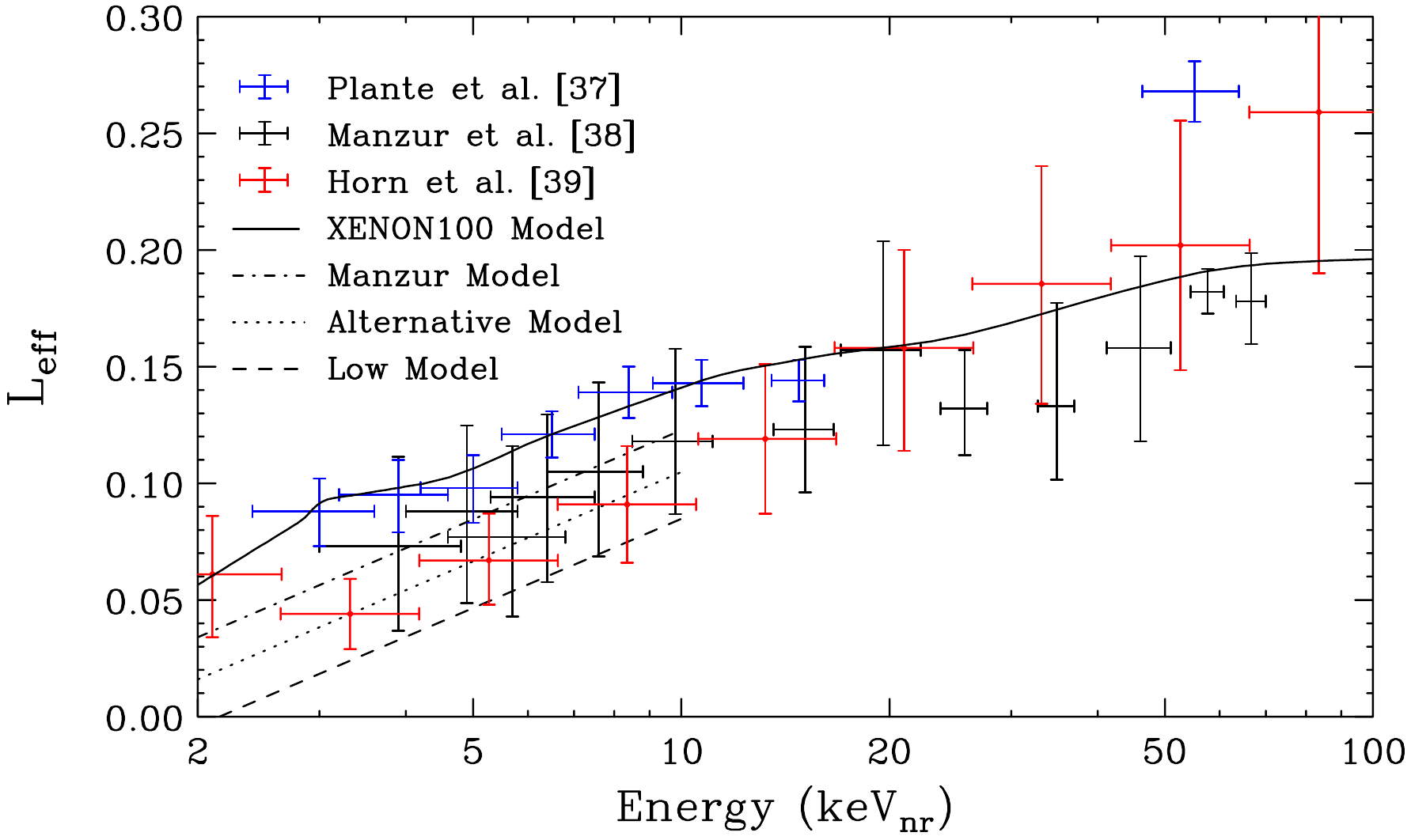}}
\caption{Recent measurements of the liquid xenon's relative scintillation efficiency, $L_{\rm eff}$~\cite{Plante:2011hw,Manzur:2009hp,Horn:2011wz}. The solid black curve denotes the values adopted by the XENON100 collaboration in their most recent analysis~\cite{global,Aprile:2013teh}. The dashed, dotted and dot-dashed curves are other $L_{\rm eff}$ models that we will consider in Secs.~\ref{events} and~\ref{uncertain}.}
\label{Leff}
\end{figure}

Two phase liquid xenon dark matter detectors such as XENON100 measure nuclear recoil events through a combination of scintillation light and ionization. The mean scintillation signal (in units of photoelectrons, PE) from a nuclear recoil of energy, $E_{\rm nr}$, is given by:
\begin{equation}
{\rm S}1=E_{\rm nr} \, L_y \, L_{\rm eff}(E_{\rm nr}) \, \frac{S_{\rm nr}}{S_{\rm ee}},
\label{s1}
\end{equation}
where $L_y$ is the light yield in photoelectrons per unit energy (at the appropriate drift field), and $L_{\rm eff}$ is the scintillation efficiency of nuclear recoil events in liquid xenon relative to that of 122 keV$_{\rm ee}$ electron recoils (see Fig.~\ref{Leff}). The quantities $S_{\rm nr}$ and $S_{\rm ee}$ account for the suppression of the scintillation signal resulting from the experiment's electric field, for nuclear and electronic recoils, respectively. The XENON100 collaboration takes these quantities to be $S_{\rm nr}=$0.95 and $S_{\rm ee}=0.58$, for their drift field of 0.53 kV/cm, and assumes that they are independent of energy (we will return to this assumption in Sec.~\ref{uncertain}). XENON100's light yield at 122 keV$_{\rm ee}$ is taken to be $L_y=2.28\pm0.04$ PE/keV$_{\rm ee}$, based on an interpolation of measurements made at 40, 80, 164, and 662 keV$_{\rm ee}$~\cite{long,Aprile:2011dd}.

In addition to scintillation light, the drift field of the XENON100 experiment allows for the observation of electrons which are ionized as the result of nuclear or electronic recoils. The mean ionization signal resulting from a nuclear recoil of energy, $E_{\rm nr}$, is given by:
\begin{equation}
{\rm S}2=E_{\rm nr} \, Q_y(E) \, Y,
\end{equation}
where $Q_y$ is the charge yield (the number of free electrons per unit energy), and $Y$ is the secondary amplification factor, or the ratio of S2 photoelectrons observed to electrons produced. The XENON100 collaboration quotes a measurement of $Y=19.5\pm0.1$ photoelectrons per electron, with fluctuations fit to a Gaussian distribution of width $\sigma_Y=6.7$ photoelectrons per electron.

The basic strategy employed in past dual phase xenon-based searches has been to use the S1 signal to approximately determine the energy of a given nuclear recoil event, and then to use the ratio of S2 and S1 signals to distinguish nuclear recoil events from electron recoil backgrounds (the ratio of S2 to S1 is significantly larger for electron recoils than for nuclear recoils). For relatively heavy dark matter particles ($\gsim$~20 GeV), this strategy is straightforward. For lighter dark matter particles, however, a number of subtle and potentially significant uncertainties come into play, making robust conclusions more difficult to draw. For dark matter particles with $m_{\rm DM}\simeq$7-10 GeV, assuming a standard choice for the velocity distribution (XENON100 derives their limits assuming a standard Maxwellian distribution with $v_0=220$ km/s and $v_{\rm esc}=544$ km/s, and a local density of 0.3 GeV/cm$^3$), a large majority of nuclear recoils will impart a few keV or less, corresponding to an average S1 signal of less than 1 photoelectron. In contrast, in the analysis producing their most recent constraints, the XENON100 collaboration imposed a threshold of S1$\ge 3$ PE (and S1$>0.3$ PE in at least two coincident photomultiplier tubes). For their assumed velocity distributions and scintillation efficiency, $L_{\rm eff}(E_{\rm nr})$~\cite{global} (see Fig.~\ref{Leff}), a 7-10 GeV dark matter particle will produce {\it no} events with S1$ \ge 3$ PE unless fluctuations around the mean predicted signal are considered. In other words, {\it all} of XENON100's events from a low-mass dark matter particle represent significant upward fluctuations in the S1 signal, well above the mean given in Eq.~\ref{s1}. If we assume that these fluctuations are simply Poisson-distributed (as the XENON100 collaboration does), and include a Gaussian S1 resolution with $\sigma=0.5\sqrt{{\rm S}1 ({\rm PE})}$ PE~\cite{long}, a 7 GeV (10 GeV) dark matter particle with an elastic scattering cross section of $\sigma_{\rm SI}=2\times10^{-41}$ cm$^2$ is predicted to produce $\sim$0.0055 (0.10) events per kg-day with S1$ \ge 3$ PE, corresponding to $\simeq$40 (800) events over XENON100's 224.6 live-days of exposure (for a fiducial mass of 34 kg, and for the cuts and efficiencies described in Ref.~\cite{Aprile:2013teh}). In contrast, the XENON100 collaboration has reported only 2 events that meet these requirements. In the next section, we will consider these two events within the context of low-mass dark matter.

\section{XENON100's Two Dark Matter Candidate Events}
\label{events}

In the analysis of 224.6 live days of data, the XENON100 collaboration identified 2 events which meet their pre-defined requirements for nuclear recoil candidates~\cite{Aprile:2012nq}. Considering the entire nuclear recoil region, the observation of a pair of events is entirely consistent with their total estimated background, $\rm{BG}_{\rm tot}=1.0 \pm 0.2$. We will argue, however, that with a closer look at the predicted distributions of this background, it is not consistent with the characteristics of the two observed events.

\begin{figure*}[t]
\centering
{\includegraphics[angle=0.0,width=3.5in]{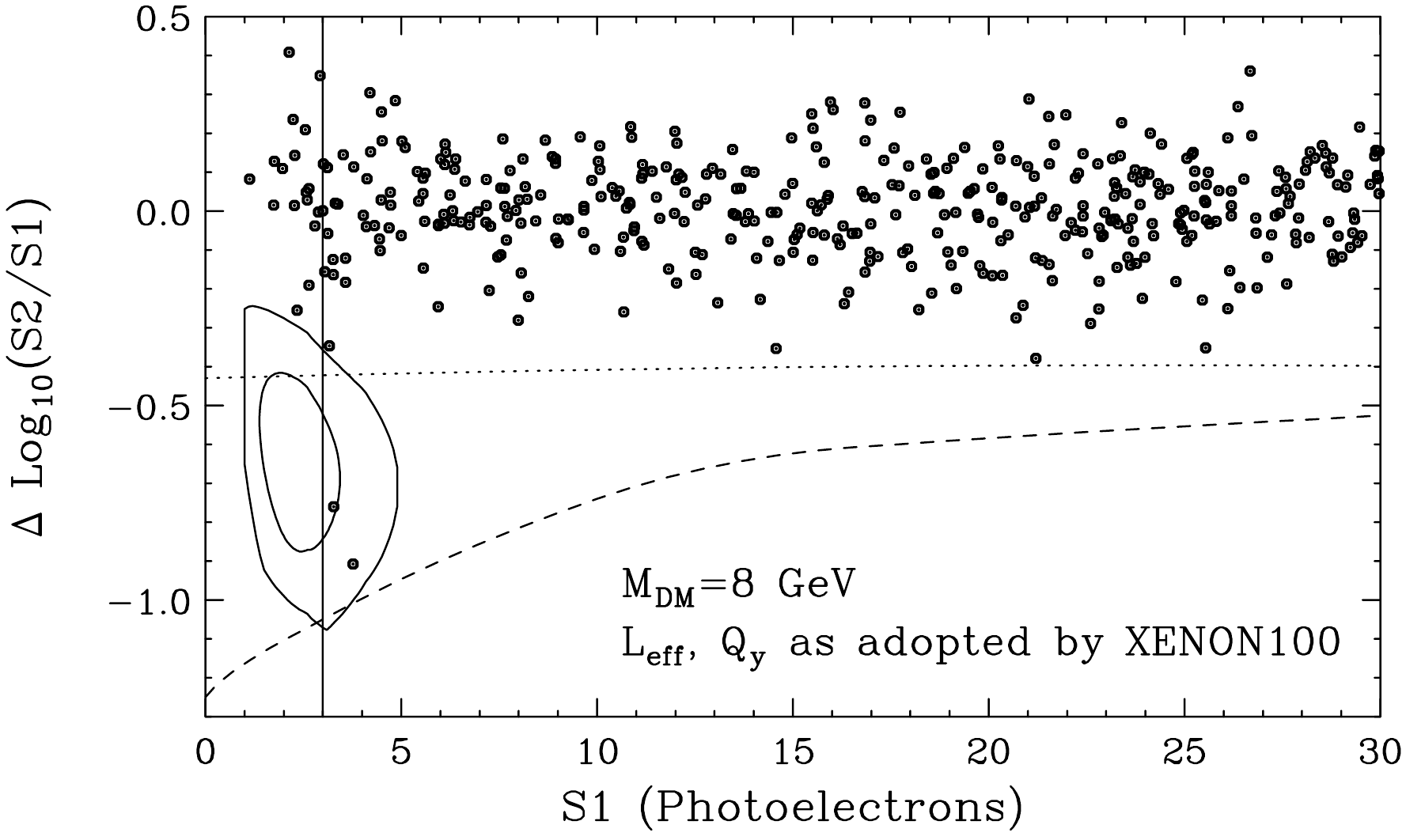}}
{\includegraphics[angle=0.0,width=3.5in]{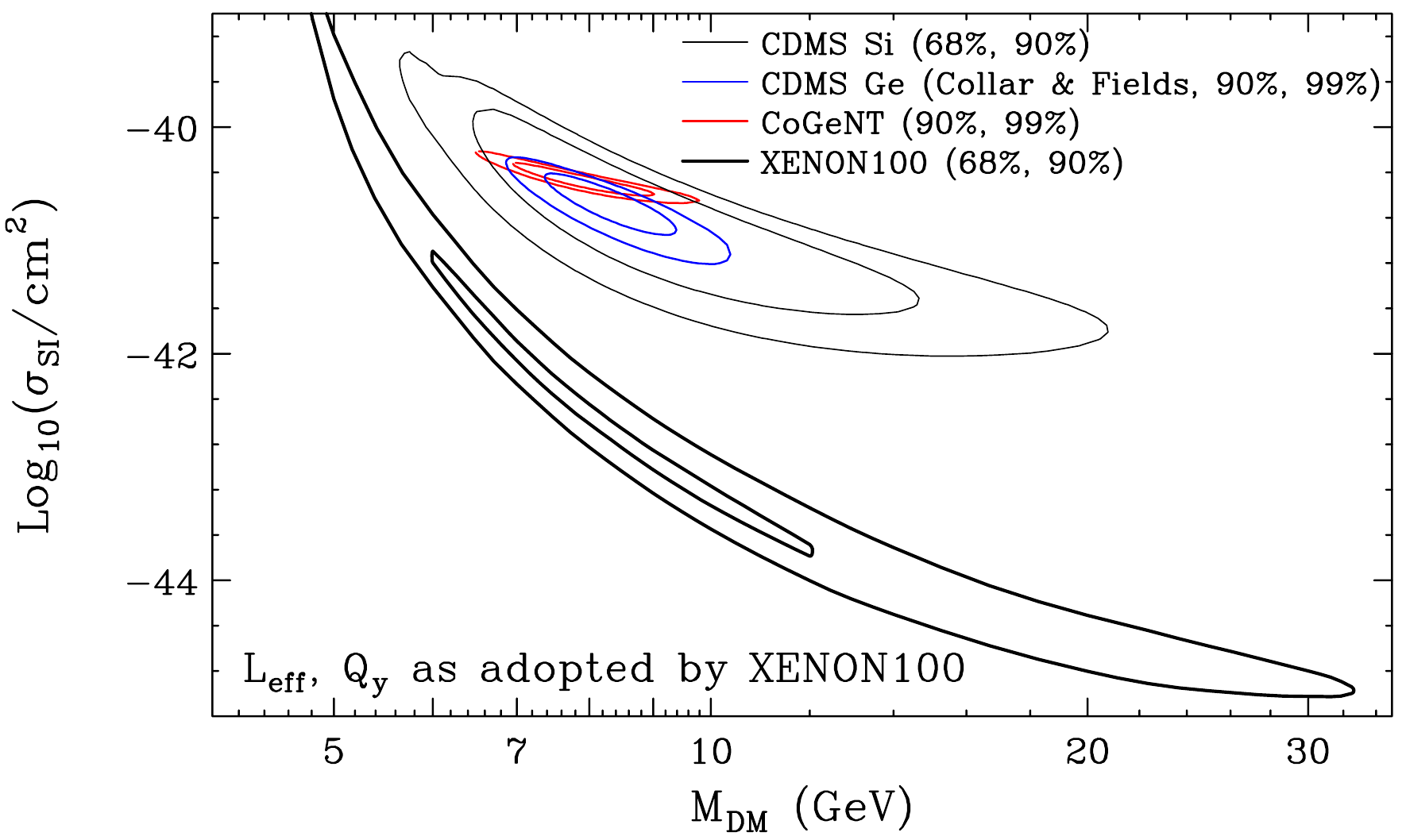}}\\
{\includegraphics[angle=0.0,width=3.5in]{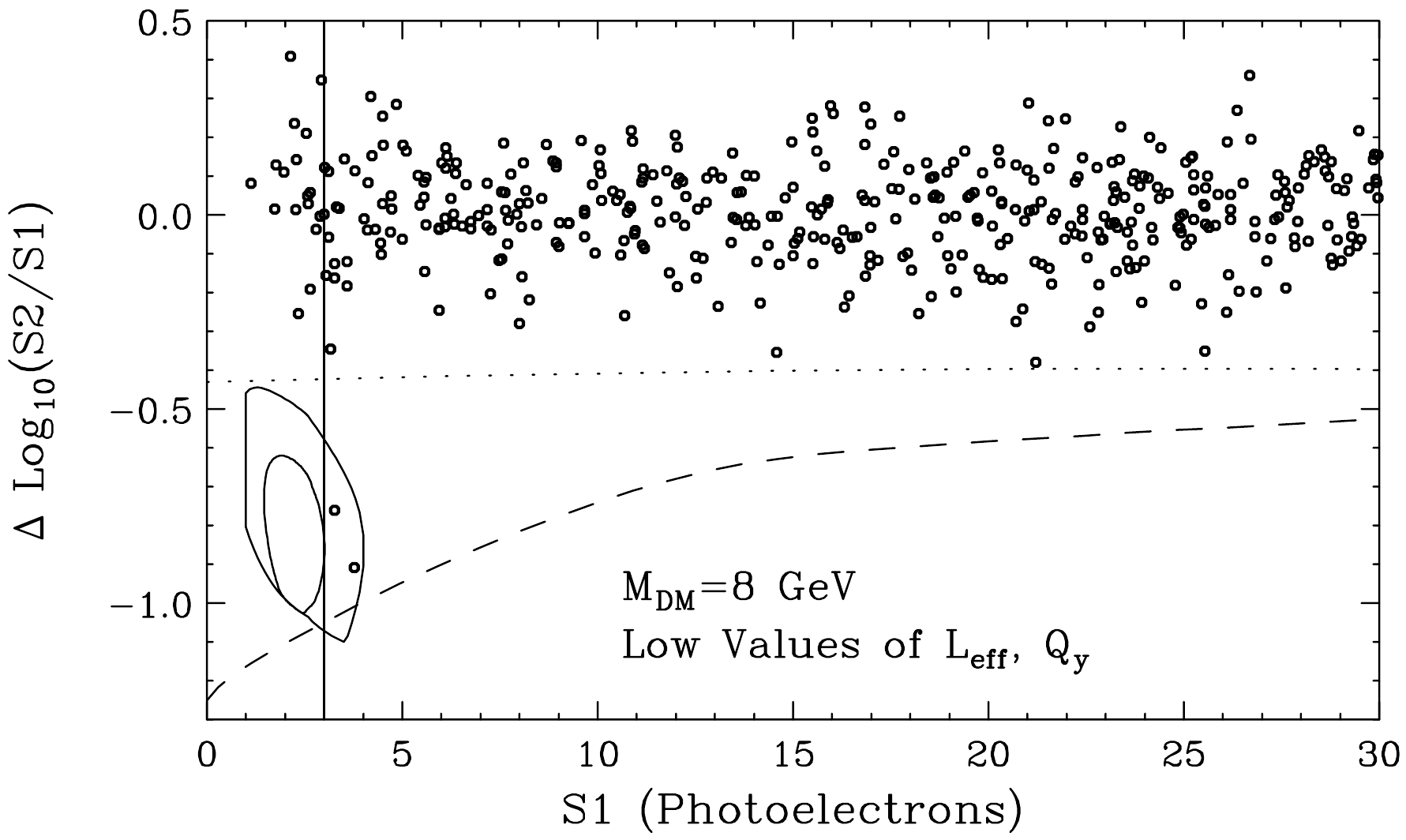}}
{\includegraphics[angle=0.0,width=3.5in]{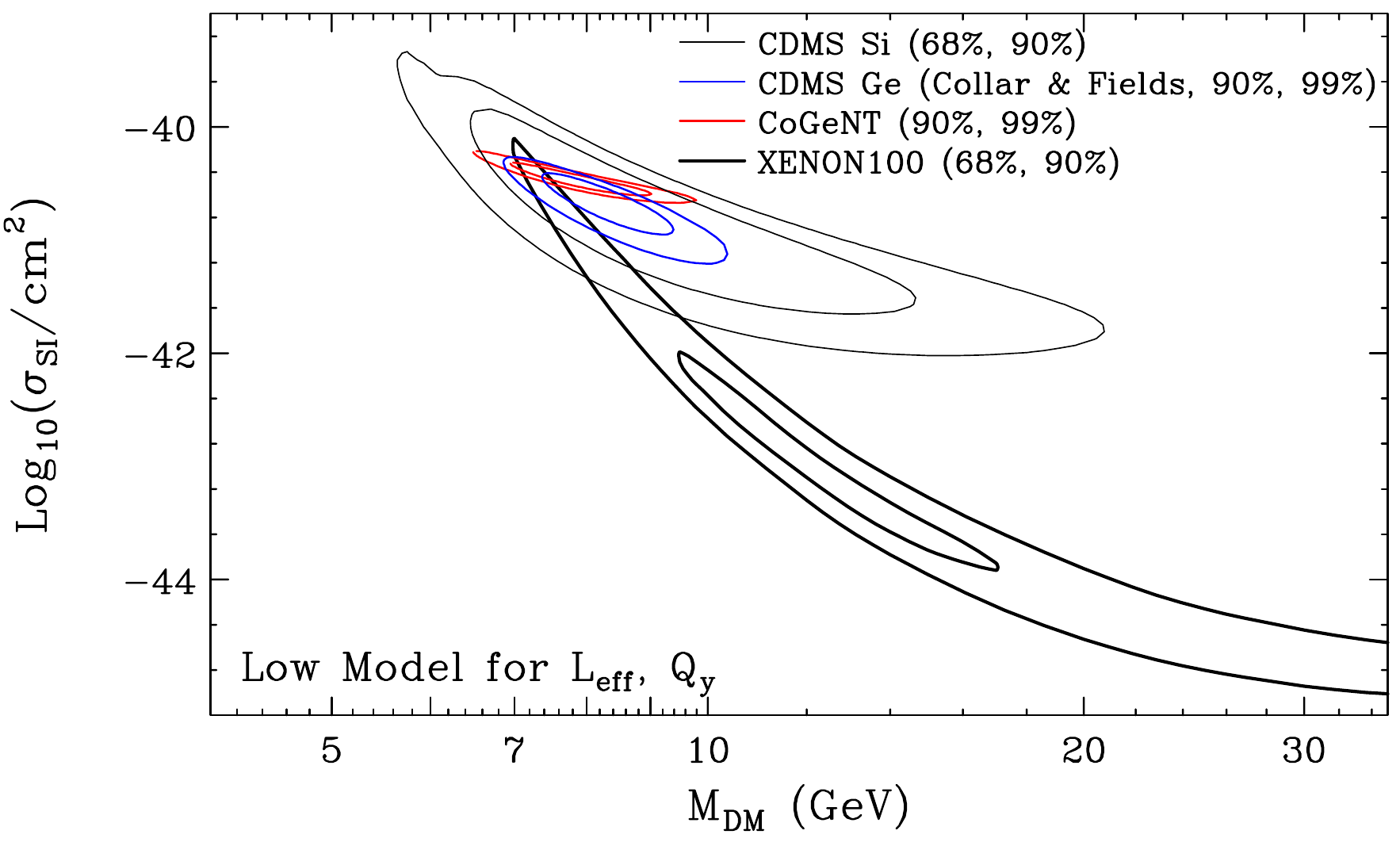}}
\caption{Left frames: The distribution of events as reported by XENON100. The closed contours represent the regions in which 50\% and 90\% of events from an 8 GeV dark matter particle are predicted to be observed. Right frames: Regions of dark matter parameter space which can account for the two events observed in XENON100's signal region (assuming that both events are produced by dark matter interactions). Lower Frames: As above, but for the low values of $L_{\rm eff}$ as shown as a dashed line in Fig.~\ref{Leff}. See text for details.}
\label{regions}
\end{figure*}

The majority of XENON100's background estimate consists of electron recoils, $\rm{BG}_{\rm ER}=0.79\pm0.16$. The 2 observed events each exhibit very low ionization-to-scintillation ratios (S2/S1), however, quite unlike electron recoil events. More specifically, these 2 events each fall slightly above XENON100's S1 threshold (S1$\simeq3-4$ PE), and have an observed ratio of S2 to S1 signals that is $\simeq$~6-8 times lower than the average electron recoil. In the left frames of Fig.~\ref{regions} we show the distribution of events reported by XENON100 in the S2/S2 vs S1 plane (as is conventional, $\Delta \log_{10} (S2/S1)$ denotes the the difference between the measured value of $\log_{10} (S2/S1)$ and the average value for an electronic recoil with the same S1~\cite{xenon100}). The dotted line near $\Delta \log_{10} (S2/S1)=-0.4$ denotes the 99.75\% electron recoil rejection line~\cite{Aprile:2012nq}. The fact that the two nuclear recoil candidate events fall so far below this line suggests that they are very unlikely to be leakage from the electron recoil event population. More quantitatively, we note that in XENON100's calibration data shown in Ref.~\cite{long}, of the 24 electron recoil events below the 99.75\% electronic recoil rejection line, only 1 falls below $\Delta \log_{10} {\rm (S2/S1)}=-0.7$~\cite{long}. Based on this calibration, we predict that the number of electron recoil background events below $\Delta \log_{10} {\rm (S2/S1)}=-0.7$ should be only $\sim$0.03, far too low to account for XENON100's 2 nuclear recoil candidate events.

The remainder of XENON100's background estimate is dominated by neutrons, $\rm{BG}_{\rm n}=0.17^{+0.12}_{-0.07}$. From this estimated rate, the probability of observing 2 or more events is approximately 3.5\%. Neutron backgrounds, however, exhibit fairly flat spectra, and are thus not predicted to be concentrated near threshold. The probability that XENON100's neutron background would lead to two events, both within $\sim$1 photoelectron of their S1 threshold, should be very low. Furthermore, the 2 observed events exhibit far lower values of (S2/S1) than are found for the vast majority of events in XENON100's neutron calibration~\cite{long}. Based on the location of the $3\sigma$ lower boundary to the neutron distribution shown in Fig.~12 of Ref.~\cite{long}, we estimate that background from neutrons below $\Delta \log_{10} {\rm (S2/S1)}=-0.8$ is even lower that that from electron recoils, $\sim$0.0005.

In light of the difficulties in explaining these 2 events, it has been suggested that they might be misidentified multiple scatter events (also known as ``gamma X'' events)~\cite{Aprile:2012nq} . As the second scatter will contribute to the S1 signal, but not to the S2, such ``false single scatters'' exhibit anomalously low values of S2/S1, just as is seen in XENON100's 2 nuclear recoil candidates. Electron recoil calibration data, however, has shown that the fraction of multiple scatter events misidentified as single scatters {\it increases} with energy~\cite{Angle:2009xb,sorensentalk}, inconsistent with the low energies of the two events in question.

Summarizing the past few paragraphs, each of the proposed backgrounds appears to be unlikely to account for XENON100's two nuclear recoil candidate events. With this in mind, we will next attempt to address whether dark matter in the parameter space favored by CoGeNT and CDMS could be responsible for these two events. To do this, we have written a Monte Carlo which, for a given dark matter mass, velocity distribution, and choice of $L_{\rm eff}$, $Q(y)$, etc., predicts the distribution of events to be observed by XENON100, in the (S2/S1) vs S1 plane. For an 8 GeV dark matter particle, and the same set of assumptions ($L_{\rm eff}$, $Q_y$, $S_{\rm nr}$, etc.) made by the XENON100 collaboration, we show this distribution in the upper left frame of Fig.~\ref{regions}. Here, the closed solid contours denote the regions of the plane in which 50\% or 90\% of the dark matter induced events are predicted to fall. Comparing this result directly to that of the XENON100 collaboration~\cite{Aprile:2013teh}, we find good agreement.

As was pointed out in Ref.~\cite{Sorensen:2012ts}, and confirmed by the XENON100 collaboration in Ref.~\cite{Aprile:2013teh}, low-mass ($\simeq$7-10 GeV) dark matter particles are predicted to produce a distribution of events that is centered around values of $\Delta \log_{10}(S2/S1)\simeq -0.75$, well below the central value predicted for heavier dark matter particles, $\Delta \log_{10}(S2/S1)\simeq -0.4$. This is due to S1 fluctuations, as discussed in Sec.~\ref{setup}; for low-mass dark matter particles, all of the events above XENON100's threshold are significant upward fluctuations in S1, and thus those nuclear recoils consistently exhibit lower than average ratios of S2-to-S1. So while XENON100's 2 recoil candidate events are near the outer edge of the (S2/S1) range predicted for heavier particles (the dashed line denotes the lower 3$\sigma$ boundary of the nuclear recoil band, applicable for $m_{\rm DM}\gsim20-30$ GeV), both of these events lie within the $\sim$1$\sigma$ range predicted for a $\sim$7-10 GeV particle.  

But although the (S2/S1)-S1 distribution predicted for a low-mass dark matter particle is in good agreement with XENON100's 2 nuclear recoil candidates, the overall number of events is not; at least under the assumptions made by the XENON100 collaboration. In particular, for the values of $L_{\rm eff}$, $Q_y$, and $S_{\rm nr}$ adopted by the XENON100 collaboration, we find that for an elastic scattering cross section of $\sigma_{\rm SI}=2\times 10^{-41}$ cm$^2$, 7, 8, or 10 GeV dark matter particles would be predicted to have produced 42, 153, or 801 events in XENON100's last run, respectively. Thus, under these assumptions, the region of dark matter parameter space collectively favored by CoGeNT and CDMS appears to predict far more events than the 2 that were observed. This can be seen in the upper right frame of Fig.~\ref{regions}, for which the region that can account for XENON100's 2 events favors significantly lower cross sections than are implied by dark matter interpretations of the CoGeNT and CDMS excesses. 

Before dismissing the CoGeNT and CDMS signals, however, one should keep in mind the considerable challenges and uncertainties involved in predicting the response of XENON100 to very low-energy nuclear recoil events. Although we will discuss these uncertainties in more detail in the following section, we will first present here a simple example of how the results at XENON100 might be reconciled with those of CoGeNT and CDMS. In particular, motivated by the experimental challenges involved in measuring liquid xenon's scintillation efficiency, $L_{\rm eff}$, and by the significant scatter between different groups' measurements of this quantity, we consider a lower value for this quantity, as shown as a dashed line in Fig.~\ref{Leff}. This low model for $L_{\rm eff}$ roughly corresponds to the 1$\sigma$ lower value of the measurements by the Yale (Manzur {\it et al}.)~\cite{Manzur:2009hp} and ZEPLIN-III (Horn {\it et al.})~\cite{Horn:2011wz} groups, but is in significant tension (we estimate roughly $\sim$4$\sigma$) with the measurements reported by members of the XENON100 collaboration~\cite{Plante:2011hw}.

Throughout this paper, for any given choice of $L_{\rm eff}$, we will adopt a model for $Q_y$ which predicts the same ratio of S2-to-S1 for any given value of S1 (neglecting fluctuations) as the original XENON100 model. This choice insures that our combined choices of $L_{\rm eff}$, $S_{\rm nr}$, and $Q_y$ will be consistent with calibration data. Note that in Fig.~\ref{Leff} we only plot the curve for our low $L_{\rm eff}$ model up to 10 keV$_{\rm nr}$ because low-mass dark matter particles are not sensitive to the high energy behavior of $L_{\rm eff}$. Quantitatively, for this choice of $L_{\rm eff}$ and other assumptions, we find that 90\% of XENON100's events from an 8 GeV dark matter particle are predicted to result from recoils in the range of 3.0 to 5.3 keV$_{\rm nr}$. 

In the lower frames of Fig.~\ref{regions}, we show the distribution of events and favored parameter space regions which result under the assumption of our low $L_{\rm eff}$ model. The predicted distribution of events in the (S2/S1) vs S1 plane changes only modestly, and is still in good agreement with XENON100's 2 observed events. The slight shift of this region toward lower values of S1 leads to a mild preference for somewhat higher dark matter masses, however. In the lower right frame, we see that in addition to the modest shift toward higher masses, this choice of $L_{\rm eff}$ has also increased the required cross section quite significantly. In particular, the regions of parameter space that are able to account for XENON100's two events now overlap (at the 90\% confidence level) with the those favored by CoGeNT and CDMS.

This exercise has demonstrated that for a low choice of $L_{\rm eff}$ (comparable to the dashed line shown in Fig.~\ref{Leff}), the results of XENON100 can be brought into consistency with the region favored by CoGeNT and CDMS. Furthermore, the two nuclear recoil candidate events reported by XENON100, which appear to be difficult to be accounted for by known backgrounds, are consistent with arising from the same dark matter particle implied by these other experiments. As stated previously, however, this choice of $L_{\rm eff}$ is in significant tension with the measurement of Ref.~\cite{Plante:2011hw} (although is consistent with those of Refs.~\cite{Manzur:2009hp,Horn:2011wz}). Ideally, we would like to find a way to reconcile XENON100 with CoGeNT and CDMS without resorting to such a low value of this quantity. In the next section, we discuss $L_{\rm eff}$ in more detail, as well as uncertainties associated with XENON100's electric field, fluctuations in the scintillation signal, and the dark matter velocity distribution. When factors such as these are considered together, we find that consistency between the results of XENON100, CoGeNT and CDMS can be obtained for somewhat higher values of $L_{\rm eff}$ than the ``low model'' considered in this section.

\section{Uncertainties In XENON100's Sensitivity to Low-Mass Dark Matter Particles}
\label{uncertain}

In this section, we consider a number of potentially relevant sources of uncertainty regarding XENON100's response to low-energy ($\sim$3-5 keV) nuclear recoils, and to low-mass dark matter particles. Among others, we focus on the uncertainties associated with the relative scintillation efficiency of liquid xenon, and with the energy dependence of the suppression of the scintillation signal resulting from XENON100's electric field. 

\subsection{The Relative Scintillation Efficiency, $L_{\rm eff}$}

We begin this section by discussing the measurements of the relative scintillation efficiency of liquid xenon to nuclear recoils, $L_{\rm eff}$. This quantity is defined as the ratio of the mean S1 signal per nuclear recoil energy, to the mean S1 signal per electron energy of a 122 electronic recoil, all at zero electric field:
\begin{equation}
L_{\rm eff}(E_{\rm nr}) \equiv \frac{S1(E_{\rm nr})/E_{\rm nr}}{S1(122 \, \rm{keV}_{\rm ee})/122\,{\rm keV}_{\rm ee}}.
\end{equation}
This quantity is conventionally defined relative to 122 keV electron recoils due to the utility of the gamma-ray line of that energy emitted by $^{57}$Co. While Lindhard theory appears to be capable of accommodating the observed behavior of $L_{\rm eff}$~\cite{Sorensen:2011bd}, and progress has been made in modeling the response of liquid xenon to nuclear recoils (such as the Noble Element Simulation Technique, NEST)~\cite{Szydagis:2011tk}, the current state of our theoretical understanding does not enable us to make detailed predictions of this quantity.

A number of groups have performed measurements of $L_{\rm eff}$, including the relatively recent measurements described in Refs.~\cite{Plante:2011hw,Manzur:2009hp,Horn:2011wz} (see Fig.~\ref{Leff}). The measurements by the Columbia group (consisting of members of the XENON100 Collaboration)~\cite{Plante:2011hw} and the Yale group~\cite{Manzur:2009hp} were each conducted using approximately monoenergetic neutron sources and were carried out in dedicated calibration detectors at zero electric field. In such a setup, the deflection angle of the neutron is used to measure the energy of the recoil, which is combined with the measured scintillation signal to provide a determination of $L_{\rm eff}$. These measurements, as shown in Fig.~\ref{Leff}, are broadly consistent with each other, although the values presented in Ref.~\cite{Manzur:2009hp} are somewhat lower than those found by Plante {\it et al}. The ZEPLIN III collaboration's measurement~\cite{Horn:2011wz}, which was conducted {\it in situ} using the ZEPLIN III detector (at non-zero field), produced values of $L_{\rm eff}$ which are similar to, but slightly lower than, those presented in Ref.~\cite{Manzur:2009hp} (and are significantly lower than those presented in Ref.~\cite{Plante:2011hw}).\footnote{In Fig.~\ref{Leff}, we show the result from Fig.~7 of Horn {\it et al.}~\cite{Horn:2011wz}, based on their S2-only event selection determination of $L_{\rm eff}$. We have chosen to show this result rather than others presented in the same paper because of their statement that this technique reduces uncertainties at low-energies associated with their Monte Carlo simulation and detection threshold efficiencies.}

While the consistency between these three measurements is difficult to rigorously evaluate (in part because Refs.~\cite{Plante:2011hw,Horn:2011wz} do not separate their quoted errors into statistical and systematic components, and because the errors are unlikely to be normally distributed) there seems to be some degree of tension between the results of these different groups. In particular, in the energy range of most interest for low-mass dark matter ($E_{\rm nr}\simeq$~3-5 keV), the results of Refs.~\cite{Plante:2011hw} and~\cite{Horn:2011wz} appear to be inconsistent at roughly the $3\sigma$ level, and the values of $L_{\rm eff}$ adopted by the XENON100 collaboration (and used in calculating the upper frames of Fig.~\ref{regions}) are not compatible with the measurements of Ref.~\cite{Horn:2011wz}. In light of this apparent tension, we think that it is reasonable to consider the possibility that the errors quoted for at least some of these measurements may be somewhat underestimated. With this in mind, we note that it has been argued that by fitting $L_{\rm eff}$ and their energy resolution independently of each other, Plante {\it et al.} find that their energy resolution (unphysically) improves in their lowest energy bins, likely leading to a systematic overestimation of $L_{\rm eff}$ at low-energies~\cite{Collar:2011wq,pcdan}. For the sake of balance, we also note that arguments have been put forth suggesting that the measurements of Manzur {\it et al.}~\cite{Manzur:2009hp} may systematically underestimate $L_{\rm eff}$~\cite{Manalaysay:2010mb}.


\begin{figure}[t]
\centering
{\includegraphics[angle=0.0,width=3.5in]{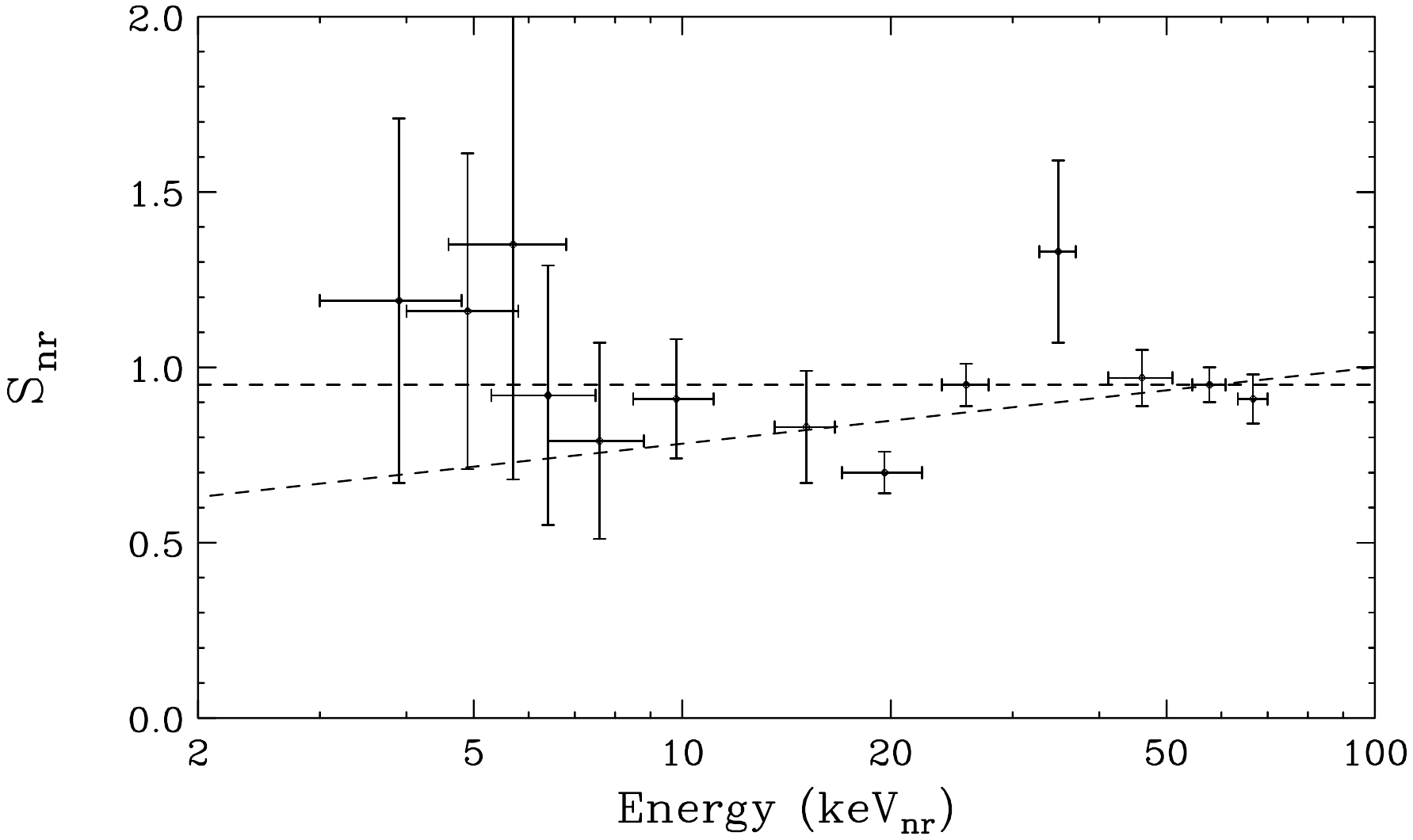}}
\caption{Measurements of $S_{\rm nr}$ as a function of recoil energy, in the presence of a 0.73 kV/cm field~\cite{Manzur:2009hp} (the closest measured value to the 0.53 kV/cm field used in XENON100). The horizontal dashed line denotes the energy-independent behavior assumed by XENON100 ($S_{\rm nr}=0.95$). In contrast, the mildly sloped dashed line provides a slightly better fit to the data.}
\label{snr}
\end{figure}

\begin{figure*}[t]
\centering
{\includegraphics[angle=0.0,width=3.5in]{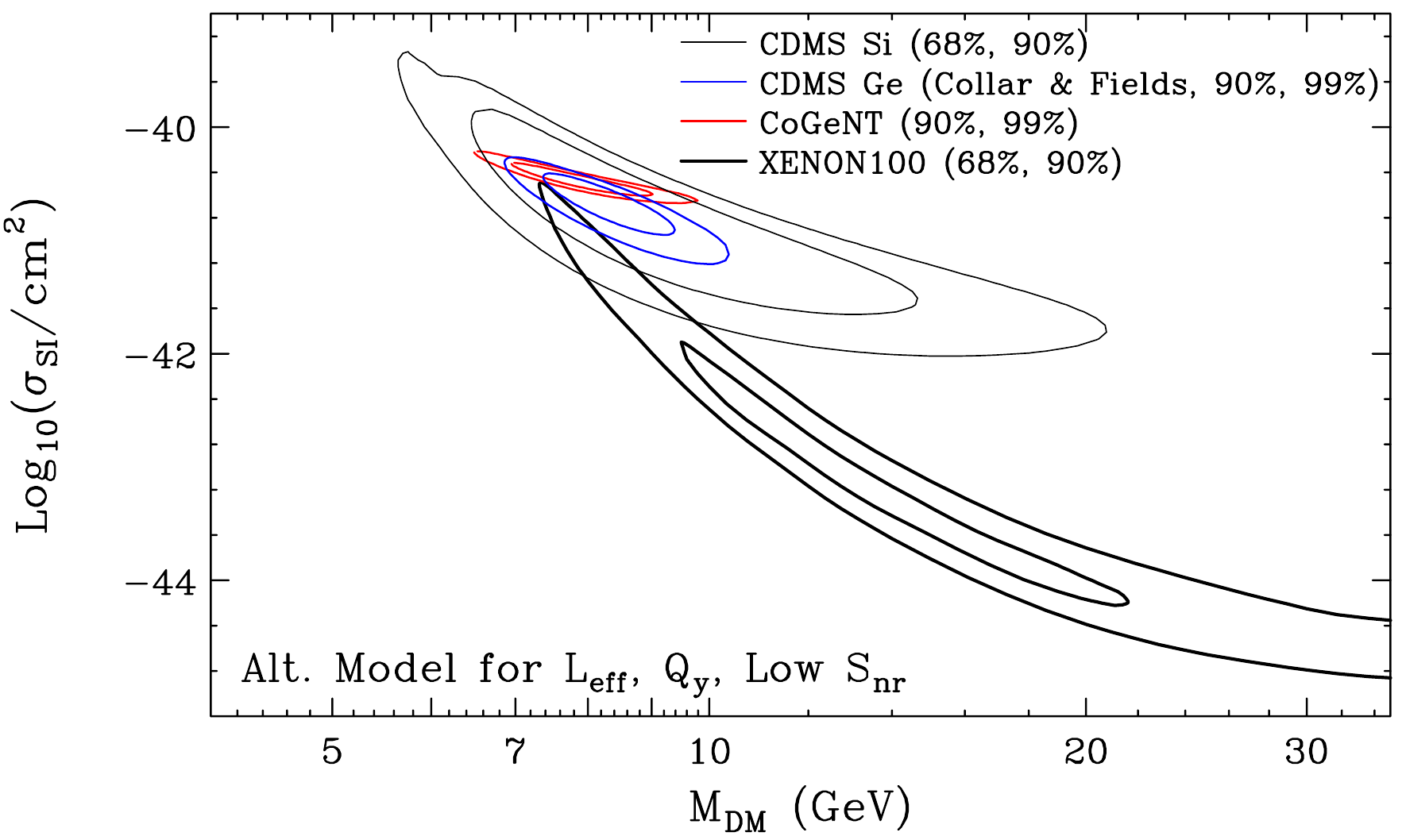}}
{\includegraphics[angle=0.0,width=3.5in]{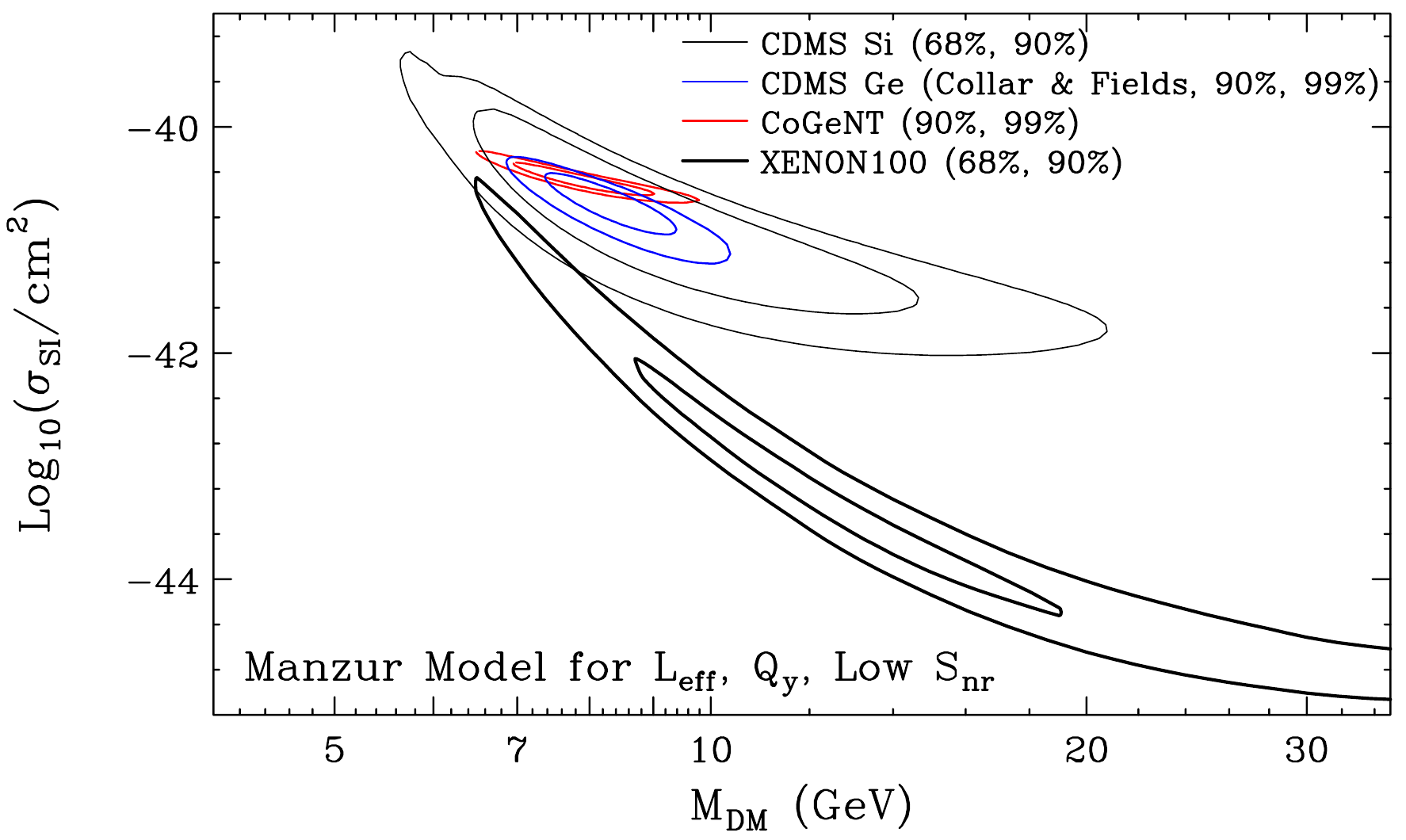}}
\caption{Regions of dark matter parameter space which can account for the two events observed in XENON100's signal region (assuming both events are produced by dark matter interactions), using the energy-dependent model for $S_{\rm nr}$ described by the sloped dashed line in Fig.~\ref{snr}. In the left and right frames, respectively, we have used the Alternative (left) and Manzur (right) models for $L_{\rm eff}$ (see Fig.~\ref{Leff}).}
\label{snrregions}
\end{figure*}

\subsection{The Impact of XENON100's Electric Field}

By definition, the quantity $L_{\rm eff}$ denotes the relative scintillation efficiency of liquid xenon at {\it zero electric field}. The electric fields used to collect and observe the ionization signal in dual-phase xenon-based detectors, however, impact the probability that a given electron will recombine with a xenon molecule, and thus alter the amount of S1 and S2 signals that result from a nuclear recoil event. The equation describing the mean S1 signal from a nuclear recoil in XENON100 (see Eq.~\ref{s1}) accounts for the effect of the electric field with the quantities $S_{\rm nr}$ and $S_{\rm ee}$, which represent suppression of the S1 signal by the electric field for nuclear and electronic recoils, respectively. XENON100's light yield, $L_y$, is also a field dependent quantity.

The XENON100 collaboration, for their drift field of 0.53 kV/cm, takes these quantities to be $S_{\rm nr}=0.95$ and $S_{\rm ee}=0.58$. These values are based on measurements of 56 keV nuclear recoils, and 122 keV electron recoils, respectively, and are explicitly assumed to be independent of energy~\cite{Aprile:2006kx}. In the case of $S_{\rm ee}$, the actual energy dependence in this quantity is absorbed into the definition of the light yield, $L_y$. Any energy dependence in $S_{\rm nr}$ relative to the value measured at 56 keV, however, will impact the interpretation of XENON100's events. 

Although there is currently no significant evidence for an energy dependence of $S_{\rm nr}$, the related uncertainties and quoted errors are large~\cite{Manzur:2009hp}, leaving open the possibility that $S_{\rm nr}$ may be smaller than assumed for low-energy recoils. The LUX Collaboration, for example, considers it likely that $S_{\rm nr}$ is energy-dependent, and have projected their sensitivities under the assumption that $S_{\rm nr}$ is significantly ($\sim$20\%) lower at keV-scale energies than at the higher energies used by XENON100 to estimate this quantity~\cite{mattidm,pcdan}.

In Fig.~\ref{snr}, we show $S_{\rm nr}$ as measured in Ref.~\cite{Manzur:2009hp}, for the case of a 0.73 kV/cm electric field (of the field strengths considered in Ref.~\cite{Manzur:2009hp}, this is the closest to XENON100's value of 0.53 kV/cm). The horizontal dashed line represents the energy-independent value of 0.95 adopted by the XENON100 collaboration. In contrast, the sloped dashed line provides a slightly better fit to the data and is similar to the model favored by the LUX collaboration. In Fig.~\ref{snrregions}, we show that by using this choice of $S_{\rm nr}$, and a model of $L_{\rm eff}$ near the central values of Horn {\it et al.}~\cite{Horn:2011wz} (the ``Alternative Model'', or dotted line in Fig.~\ref{Leff}) or Manzur {\it et al.}~\cite{Manzur:2009hp} (the ``Manzur Model'', or dot-dashed line in Fig.~\ref{Leff}), we can find consistency (or near consistency in the case of the Manzur Model) between the results of XENON100, CoGeNT, and CDMS. Note that, in contrast to Refs.~\cite{Plante:2011hw,Manzur:2009hp},  Horn {\it et al.}~\cite{Horn:2011wz} measured $L_{\rm eff}$ in the presence of an electric field, and thus have also implicitly measured the energy dependance of $S_{\rm nr}$.

Efforts are currently underway to measurements $S_{\rm nr}$ (and $Q_y$) over a range of electric fields and recoil energies~\cite{fnal}. Such measurements will be essential to interpreting low-energy nuclear recoil events in liquid xenon detectors.

\subsection{Low-Energy Efficiencies, S1 Fluctuations, and Other Considerations Near Threshold}

As discussed in Sec.~\ref{setup}, XENON100's sensitivity to dark matter particles lighter than $\sim$10 GeV is entirely reliant on the small fraction of the highest energy recoil events which produce S1 signals that are well above the mean value predicted ({\it ie.} upward fluctuations from the mean S1 signal described by Eq.~\ref{s1}). In this respect, XENON100 can only observe events which are on the tail of the recoil energy distribution {\it and} on the tail of the distribution of S1 PMT fluctuations. The XENON100 collaboration treats the distribution of their S1 fluctuations as Poissonian. In actuality, such fluctuations are unlikely to be so simple. For example, the LUX collaboration's Monte Carlo simulation accounts for many sources of stochastic fluctuations, including those from light collection, quantum efficiency, recombination, the Fano factor, excitation vs. ionization channels, dE/dx, and particle track history, etc~\cite{mattseminar,pcmatt}. While some of these variations may be well described by a Poisson distribution, others are not. If the actual distribution of S1 signals around the mean is less broad than the Poisson distribution assumed by the XENON100 collaboration, it could lead them to overestimate their sensitivity to low-energy nuclear recoils~\cite{Collar:2011wq}. As a naive example, we note that by treating these fluctuations as binomially distributed (instead of Poisson), the event rate predicted at XENON100 can be reduced by up to a factor of roughly 50\%. Without a sophisticated Monte Carlo which accounts for these many stochastic processes, we cannot reliably estimate the impact of any non-Poissonian fluctuations in the S1 signal of low-energy nuclear recoils.

We also remind the reader that the overall efficiencies of XENON100 are very sensitive to the precise value of S1 in the range of nuclear recoil energies relevant for low-mass dark matter particles (see Fig.~1 of Ref.~\cite{Aprile:2013teh}). Even very small changes in the efficiency curve could significantly alter the rate of nuclear recoil events predicted from low-mass dark matter particles.

\begin{figure*}[t]
\centering
{\includegraphics[angle=0.0,width=3.5in]{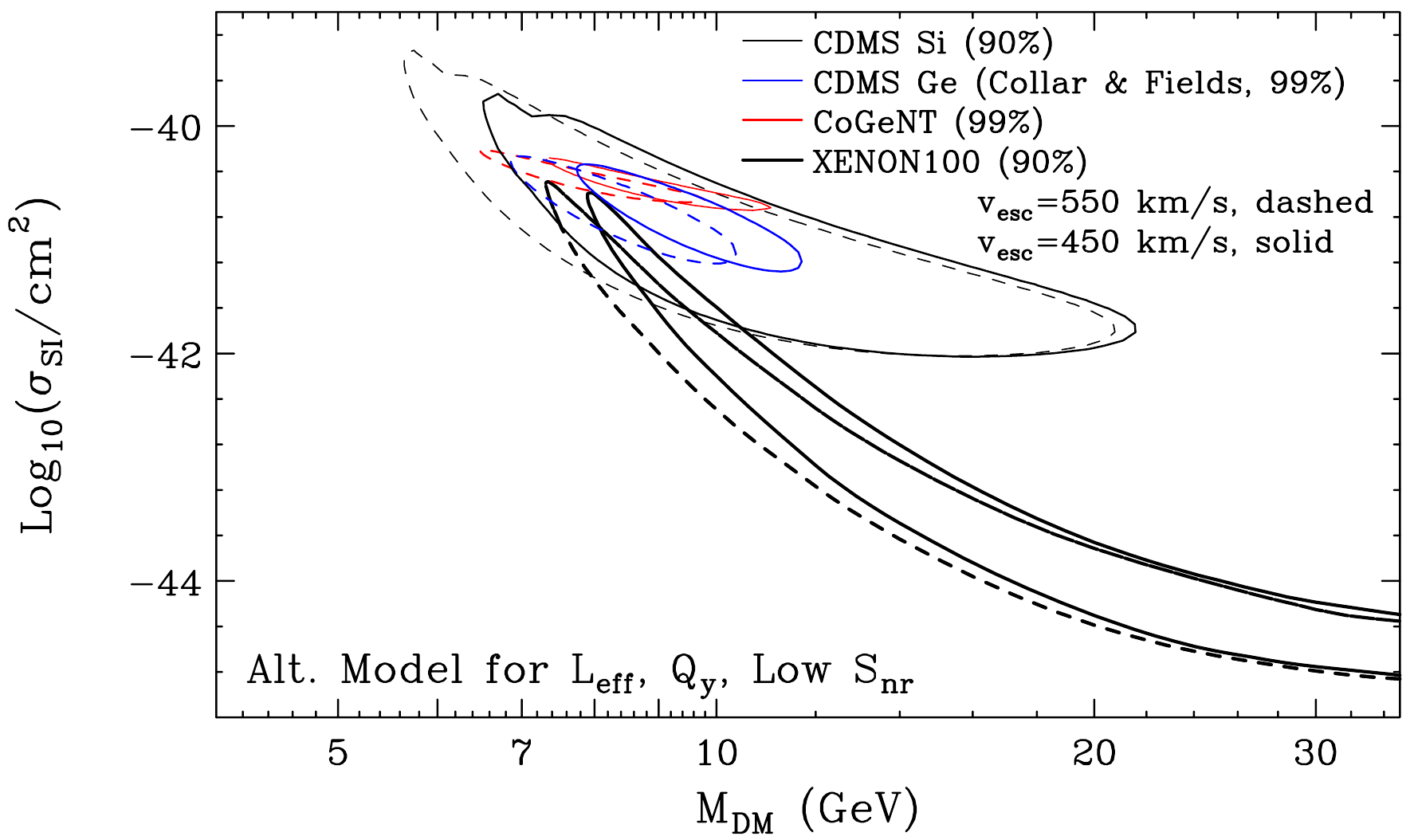}}
{\includegraphics[angle=0.0,width=3.5in]{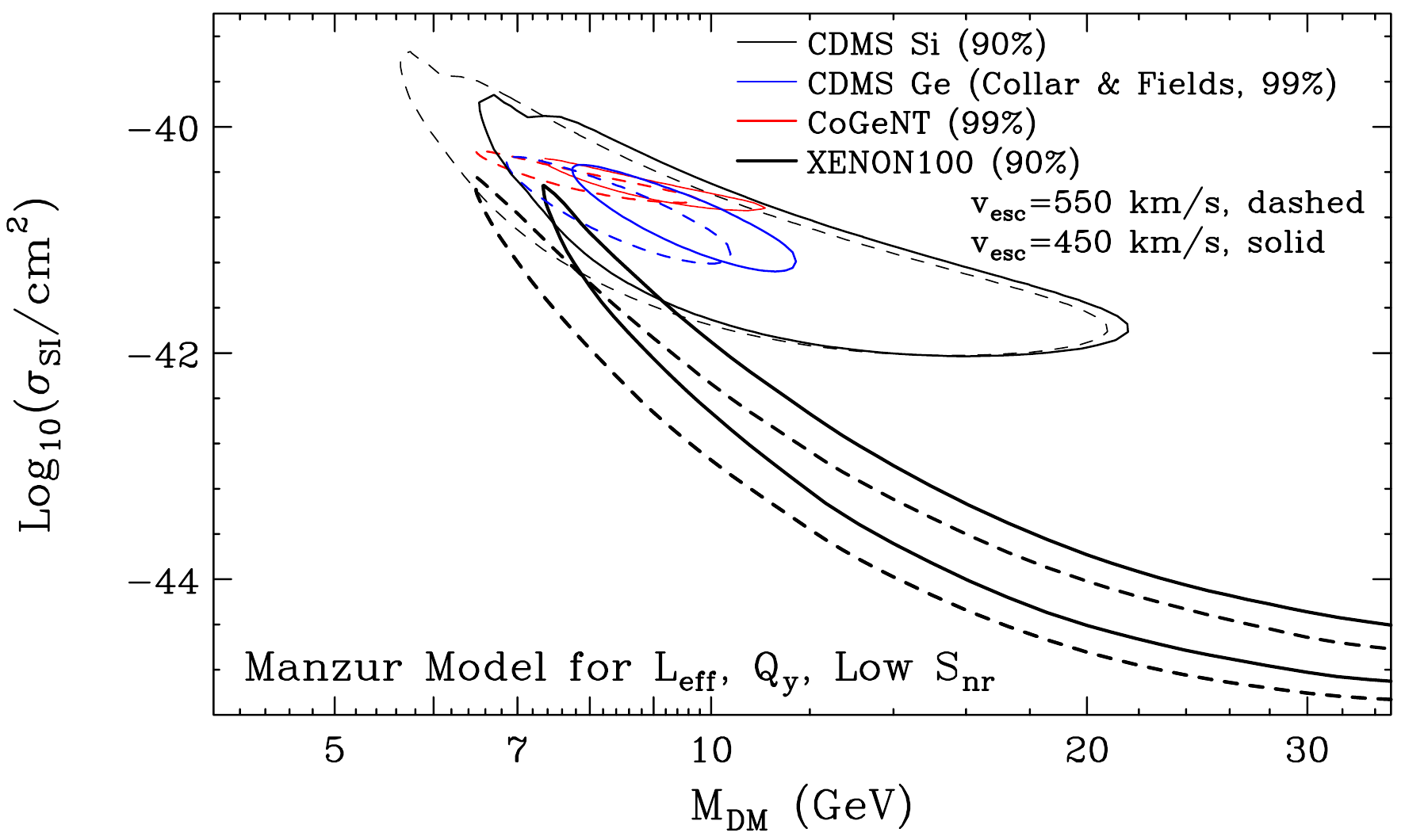}}
\caption{As in Fig.~\ref{snrregions}, but for two choices of the Galactic escape velocity, $v_{\rm esc}$. We thank Julien Billard of the CDMS Collaboration and Nicole Fields of CoGeNT, for providing the CDMS and CoGeNT contours for $v_{\rm esc}=450$ km/s.}
\label{vesc}
\end{figure*}

\subsection{The Dark Matter Velocity Distribution}

So far, we have restricted ourselves to discussing uncertainties involved with the response of the XENON100 detector itself. Variations in the assumptions regarding the velocity distribution of the dark matter~\cite{Mao:2013nda,v5,v2,v1} or its interactions~\cite{Frandsen:2013cna,Feng:2011vu} could also help to reconcile their constraint with the signals reported by CoGeNT and CDMS. In this subsection, we consider the former of these possibilities.

As is conventional, the XENON100 collaboration has adopted a maxwellian velocity distribution for the dark matter, with a local circular velocity of $v_0=220$ km/s and a galactic escape velocity of $v_{\rm esc}=544$ km/s~\cite{v4}. The precise value of the escape velocity can be important when considering direct detection signals appearing very near experimental thresholds. In particular, we find that for an 8 GeV dark matter particle, and our low $L_{\rm eff}$ model, lowering $v_{\rm esc}$ from 544 km/s to 500 km/s (450 km/s) reduces the overall event rate predicted for XENON100 by a factor of 1.55 (3.3). In contrast, CoGeNT's signal extends well above their energy threshold, making their signal less sensitive to the escape velocity assumed. In Fig.~\ref{vesc}, we show the impact of this parameter on the favored dark matter parameter space.  A low value of the escape velocity can mildly help to reconcile XENON100 with CoGeNT and CDMS. We also direct the reader to Ref.~\cite{Mao:2013nda}, which considers a range of cosmologically motivated dark matter velocity distribution models, demonstrating that the tension between the results of XENON100 and CDMS can be significantly reduced relative to that found in the case of a standard Maxwellian distribution.

\section{Predictions for LUX}
\label{others}

In this paper, we have argued that there are sufficient uncertainties in the details of XENON100's response to low-energy ($\sim$3-5 keV) nuclear recoils that it is possible that the results of their analysis of 224.6 live days of data may be consistent with the regions of dark matter parameter space favored by CoGeNT and CDMS. Furthermore, we have argued that the two nuclear recoil events reported by XENON100 are not easily accounted for with published backgrounds, but exhibit the characteristics (S1 and S2/S1) predicted for a dark matter particle in the mass range favored by CoGeNT and CDMS.  If these two events arise from the same dark matter particle being observed by CoGeNT and CDMS, then the upcoming LUX experiment~\cite{Akerib:2012ys} should detect a significant excess of low-energy nuclear recoil events (as should XENON1T).

For the purposes of detecting low-mass dark matter particles, the LUX experiment improves on XENON100 in two important respects. Firstly, their fiducial mass of 100 kg is a factor of almost three time larger than XENON100's. Even more important in the case of low-mass particles is LUX's much higher light yield ($L_y$), which has been measured to be at least 2, or perhaps 3, times as high as XENON100's~\cite{Fiorucci:2013yw,Akerib:2012ak}. For dark matter particles with a mass very close to XENON100's S1 threshold ($\sim$7-10 GeV), we find that increasing the light yield by a factor of 2 (3) enhances the rate of events with S1$\ge3$ PE by a factor of $\simeq$4 ($\simeq$7-9). For heavier particles, the impact of the light yield on the event rate is much more modest. Accounting for LUX's larger target mass and greater light yield, we predict that a 7-10 GeV dark matter particle that produces an expectation value of 2 events over 224.6 days at XENON100 will generate an average of 3.1-7.6 events per month at LUX (assuming similar cuts and efficiencies for LUX as for XENON100). Combining this with the 90\% Poisson uncertainty on the underlying rate implied by XENON100's two events, and with the rates observed by CoGeNT and CDMS, we predict that if XENON100's two events are from the same 7-10 GeV dark matter particles responsible for the CoGeNT and CDMS signals, then LUX should observe between 3 and 24 events per month. LUX should be sensitive even to the low end of this predicted range.

We note that the projected sensitivity to low-mass dark matter particles as presented by the LUX collaboration has been quite conservative, in large part motivated by many of the same considerations discussed in this paper. For example, in Fig.~12 of Ref.~\cite{Akerib:2012ak}, the LUX collaboration claims {\it no} sensitivity to dark matter particles lighter than 7 GeV (8 GeV) for what they describe as realistic (very conservative) assumptions pertaining to their light collection (an S2-only analysis of LUX data would be sensitive to lower masses, however). Comparing this to the constraints quoted by the XENON100 collaboration, this illustrates the very significant role that uncertainties (such as those regarding $L_{\rm eff}$ and $S_{\rm nr}$) presently play in the interpretation of low-energy data from xenon-based experiments.

\section{Summary and Conclusions}
\label{summary}

Although the excesses of low-energy events observed by CoGeNT and CDMS can be interpreted as evidence for low-mass ($m_{\rm DM}\simeq$~7-10 GeV) dark matter particles, such a scenario appears to be inconsistent with the constraints published by the XENON100 collaboration. In this paper, we have revisited this constraint and discussed a number of uncertainties that could potentially help to reconcile it with the CoGeNT and CDMS signals. In particular, we have discussed uncertainties related to:
\begin{itemize}
\item{The relative scintillation efficiency of liquid xenon, $L_{\rm eff}$.}
\item{Energy-dependance in the suppression of the scintillation signal resulting from XENON100's electric field, $S_{\rm nr}$.}
\item{Non-Poissonian fluctuations of the scintillation signal around the mean, and other factors which may impact the near-threshold efficiency of XENON100.}
\item{The dark matter's velocity distribution, and in particular the escape velocity of the galaxy.}
\end{itemize}

Taken together, we find plausible scenarios in which the results of XENON100 could be consistent with a dark matter interpretation of CoGeNT and CDMS. We also point out that while the two nuclear recoil candidate events observed by XENON100 each exhibit ionization and scintillation signals consistent with resulting from 7-10 GeV dark matter particles, these events each produced too little ionization to be likely attributable to either electronic or neutron backgrounds.  In particular, based on XENON100's calibration data, we estimate that they should have observed only $\sim$~0.03 background events with so little relative ionization. Although it is difficult to entirely rule out other backgrounds, the characteristics of these two events are quite suggestive of a low-mass dark matter interpretation.

We do not feel that the conclusions reached in this paper are in radical departure from those presented by the XENON100 collaboration. While their estimated event rate from an 8 GeV dark matter particle with an elastic scattering cross section of $\sigma_{\rm SI} = 2\times10^{-41}$ is $180_{-53.3}^{+220}$~\cite{Aprile:2013teh}, and well above the 2 events observed, they also state that the errors in this estimated rate are dominated by systematics, such as those associated with $L_{\rm eft}$ and $Q_y$. If the relevant systematic errors (including also those associated with the energy dependance of $S_{\rm nr}$) are even modestly larger than estimated by the XENON100 collaboration, there could plausibly exist a low-mass region of parameter space in which the results of XENON100 are consistent with the signals reported by CoGeNT and CDMS.



\bigskip

\section*{Acknowledgements} We would like to thank Matthew Szydagis, Eric Dahl, Nicole Fields, Lauren Hsu, Rafael Lang, Dan McKinsey, Peter Sorensen, Andrew Sonnenschein, and Juan Collar for helpful discussions. We would also like to thank the XENON100, CoGeNT and CDMS collaborations for providing many of the parameter regions and data points shown in the figures throughout this paper. This work has been supported by the US Department of Energy.

\end{document}